%% file: 0_main.tex
\documentclass[conference]{IEEEtran}
\IEEEoverridecommandlockouts
\usepackage{cite}
\usepackage{amsmath,amssymb,amsfonts}
\usepackage{algorithmic}
\usepackage{graphicx}
\usepackage{subcaption}
\usepackage{flushend}
\usepackage{textcomp}
\usepackage{xcolor}
\usepackage{siunitx}
\usepackage{cleveref}
\usepackage{booktabs,siunitx,tabularx, stfloats}
\usepackage{enumitem}
\usepackage{bm}
\usepackage[font=small,skip=5pt]{caption}

\crefname{section}{Section}{Sections}
\crefname{figure}{Figure}{Figures}
\crefname{table}{Table}{Tables}

\def\BibTeX{{\rm B\kern-.05em{\sc i\kern-.025em b}\kern-.08em
    T\kern-.1667em\lower.7ex\hbox{E}\kern-.125emX}}

\usepackage{glossaries}
\newacronym{asv}{ASV}{Automatic Speaker Verification}
\newacronym{tts}{TTS}{Text to Speech}
\newacronym{cnn}{CNN}{Convolutional Neural Network}
\newacronym{roc}{ROC}{Receiver Operating Characteristic}
\newacronym{moe}{MoE}{Mixture of Experts}
\newacronym{eer}{EER}{Equal Error Rate}
\newacronym{auc}{AUC}{Area Under the Curve}

\begin{document}

\title{Leveraging Mixture of Experts \\for Improved Speech Deepfake Detection
\thanks{This material is based on research sponsored by the Defense Advanced Research Projects Agency (DARPA) and the Air Force Research Laboratory (AFRL) under agreement number FA8750-20-2-1004. The U.S. Government is authorized to reproduce and distribute reprints for Governmental purposes notwithstanding any copyright notation thereon. The views and conclusions contained herein are those of the authors and should not be interpreted as necessarily representing the official policies or endorsements, either expressed or implied, of DARPA and AFRL or the U.S. Government.
This work was supported by the FOSTERER project, funded by the Italian Ministry of Education, University, and Research within the PRIN 2022 program. This work was partially supported by the European Union under the Italian National Recovery and Resilience Plan (NRRP) of NextGenerationEU, partnership on ``Telecommunications of the Future'' (PE00000001 - program ``RESTART'').}
}

\author{\IEEEauthorblockN{
Viola Negroni,
Davide Salvi,
Alessandro Ilic Mezza,
Paolo Bestagini,
Stefano Tubaro
}\vspace{0.35em}
\IEEEauthorblockA{\textit{Dipartimento di Elettronica, Informazione e Bioingegneria (DEIB), Politecnico di Milano}\\
\{viola.negroni, davide.salvi, alessandroilic.mezza, paolo.bestagini, stefano.tubaro\}@polimi.it
}
}

\maketitle

\begin{abstract}

Speech deepfakes pose a significant threat to personal security and content authenticity.
Several detectors have been proposed in the literature, and one of the primary challenges these systems have to face is the generalization over unseen data to identify fake signals across a wide range of datasets.
In this paper, we introduce a novel approach for enhancing speech deepfake detection performance using a Mixture of Experts architecture. 
The Mixture of Experts framework is well-suited for the speech deepfake detection task due to its ability to specialize in different input types and handle data variability efficiently.
This approach offers superior generalization and adaptability to unseen data compared to traditional single models or ensemble methods.
Additionally, its modular structure supports scalable updates, making it more flexible in managing the evolving complexity of deepfake techniques while maintaining high detection accuracy. 
We propose an efficient, lightweight gating mechanism to dynamically assign expert weights for each input, optimizing detection performance.
Experimental results across multiple datasets demonstrate the effectiveness and potential of our proposed approach.


\end{abstract}


\input{1_intro}
\input{2_method}
\input{3_setup}
\input{4_results}
\input{5_conclusion}


\bibliographystyle{IEEEtran}
\bibliography{bstcontrol.bib, biblio.bib}

\end{document}

%% file: 1_intro.tex
\section{Introduction}

In recent years, methods for generating or modifying synthetic media have become increasingly widespread and accessible.
In the audio domain, the use of deepfakes involves the synthesis of speech using the voice of a target speaker, making them say arbitrary utterances.
This kind of AI-generated media pose a synthetic threat to the integrity of digital content, especially as the techniques used to generate them evolve at an alarming pace~\cite{bhagtani2024recent, patel2023deepfake}. 
To counter this threat, researchers have proposed various speech deepfake detection methods, employing a range of model architectures, processing techniques, and strategies~\cite{cuccovillo2022open, almutairi2022review}.
Likewise, extensive research has been devoted to assessing the robustness of the developed systems to assess their applicability in real-world conditions~\cite{yadav2024fairssd, salvi2023reliability, Borzi_2022_CVPR}.
Although these systems have achieved remarkable performance in controlled scenarios~\cite{yang2024robust, xie2023domain, li2023voice, mari2023all}, the challenge of effectively generalizing their capabilities to unseen data and increasingly sophisticated speech deepfakes remains critical.
This issue is further complicated by the unique characteristics of the state-of-the-art speech deepfake datasets, which include diverse generation methods and multilingual data, making it reasonable to consider them as distinct domains.
For these reasons, developing flexible tools capable of dealing with increasingly variable scenarios is becoming crucial.

To tackle this challenge, we propose a novel method based on a \gls{moe} framework.
\glspl{moe} are hierarchical models comprising an ensemble of $N$ experts governed by a probabilistic gating function~\cite{jacobs1991adaptive, yuksel2012twenty, masoudnia2012survey, cai2024survey}.
While all experts work toward the same goal, such as classification or regression, the gating function dynamically determines the contribution of each expert in inferring the final output.
In particular, the expert weights are typically calculated by applying a softmax function to the logits produced by the gating network so that each weight can be interpreted as the percentage of involvement of each expert in the decision-making process relative to the given input sample.

In the original formulation by~\cite{jacobs1991adaptive}, both the experts and the gating functions were implemented using feedforward neural networks and jointly trained to minimize a single objective function in an end-to-end fashion.
This way, the gating network proportionally distributes the gradient flow across all experts, effectively rewarding the experts with the best performance. In turn, this encourages the subdivision of the problem space into homogeneous regions~\cite{baldacchino2016moe} and localizes each expert onto a specific distribution on the input feature space via an underlying competitive learning process. Indeed, this family of \glspl{moe}, also known as mixtures of \textit{implicitly localized} experts~\cite{masoudnia2012survey}, are rooted in the idea of partitioning a complex task into sub-problems easier for each expert to learn and solve, fundamentally amounting to a divide and conquer strategy.

We adapt the \gls{moe} framework to the speech deepfake detection task by leveraging expert specialization across different domains, i.e., state-of-the-art datasets.
To the best of our knowledge, this work constitutes the first application of the \gls{moe} architecture in speech deepfake detection.
Our findings indicate that our approach shows promise in enhancing generalization and adaptability to a range of deepfake techniques, making it a valuable contribution to the speech deepfake detection domain.


%% file: 2_method.tex
\glsreset{moe}

\section{Proposed Method}
\label{sec:method}


\subsection{Problem Formulation}
\label{subsec:problem}

The speech deepfake detection problem is formally defined as follows.
Let us consider a discrete-time input speech signal $\mathbf{x}$ sampled at a frequency $f_\text{s}$ and associated with a class $y \in \{0, 1\}$, where \num{0} denotes that the signal is authentic and \num{1} indicates that it has been synthetically generated.
The goal of this task is to develop a detector $\mathcal{D}$ that estimates the class of the signal $\mathbf{x}$ as $\hat{y} \in [0,1]$, where $\hat{y}$ is 
the likelihood of the signal $\mathbf{x}$ being fake.
 
\subsection{Proposed Systems}
\label{subsec:system}

In the proposed approach, we design the detector $\mathcal{D}$ as a \gls{moe} specifically tailored for the speech deepfake detection task.
Our \gls{moe} system is composed of multiple detectors, each pre-trained on a different speech deepfake dataset, thus becoming an ``expert'' in that domain. 
We denote the $i$-th expert as $\mathcal{E}_i$, where $i \in \{1, 2, \ldots, N\}$, and $N$ represents the total number of considered domains.
While all experts are fed with the same audio signal $\mathbf{x}$, the input of the gating network $\mathcal{G}$ varies based on the chosen \gls{moe} implementation.

We consider two different architectures for the \gls{moe} framework, which we will later refer to as \textit{standard} and \textit{enhanced}, respectively. 
These are both \textit{dense} \glspl{moe}~\cite{cai2024survey}, meaning that every expert $\mathcal{E}_i$ is queried for each input $\mathbf{x}$. 

\vspace{.5em} \noindent \textbf{Standard MoE}.
The standard model is a classic \gls{moe} inspired by \cite{jacobs1991adaptive}, where both $\mathcal{G}$ and the experts $\mathcal{E}_i$ directly receive the input audio excerpt $\mathbf{x}$.
Each expert returns a vector of logits as
\begin{equation}
    \mathbf{z}_i = \mathcal{E}_i(\mathbf{x}),
\end{equation}
which could in principle be used for classification already.
The output of the gating network $\mathcal{G}$ is
\begin{equation}
    \bm{\alpha} = \mathcal{G}(\mathbf{x}),
\end{equation}
where $\bm{\alpha} \in \mathbb{R}^N$.
The $i$-th element of $\bm{\alpha}$ is denoted $\alpha_i$ and acts as a weight for the output of the $i$-th expert.
The final \gls{moe} output is obtained as
\begin{equation}
    \label{eq:moe}
    \mathbf{z} = \sum_{i=1}^N \alpha_i \cdot \mathbf{z}_i.
\end{equation}
The final prediction score $\hat{y}$ of the complete speech deepfake detector $\mathcal{D}$ is obtained by applying a softmax function to the weighted logits vector $\mathbf{z}$.


\vspace{.5em} \noindent \textbf{Enhanced MoE}.
The enhanced \gls{moe} architecture differs from the standard \gls{moe} described above in that the gating network, instead of being fed the raw audio signal $\mathbf{x}$, relies on the internal representations produced by the $N$ experts.
In this design, the gating network uses the first few layers of the experts as embedding extractors and takes as input a concatenation of the embeddings from each expert, along with a combined embedding.

Formally, each expert embedding $\mathbf{e}_\text{i}\in\mathbb{R}^d$ is extracted from the last batch normalization layer of $\mathcal{E}_i$.
The combined embedding $\mathbf{w}\in\mathbb{R}^d$, in turn, is created through a learnable weighted element-wise multiplication of $\mathbf{e}_\text{1}, ..., \mathbf{e}_\text{N}$.
Considering $N=4$ and $d=64$, the combined embedding is calculated as
\begin{equation}
\mathbf{w} = (\mathbf{e}_1 \odot \mathbf{e}_2 \odot \mathbf{e}_3 \odot \mathbf{e}_4) \odot \mathbf{p}, 
\end{equation}
where $\odot$ denotes element-wise multiplication, and $\mathbf{p} \in \mathbb{R}^{d}$ is a learnable parameter.
The final concatenated embedding vector $\mathbf{e}_{\text{input}}$ is formed by
\begin{equation}
\mathbf{e}_{\text{input}} = \left[\mathbf{e}_1^T, \mathbf{e}_2^T, \mathbf{e}_3^T, \mathbf{e}_4^T, \mathbf{w}^T\right]^T.
\label{eq:input_gating}
\end{equation}
This specific design is intended to fully leverage the domain-specific knowledge acquired during pre-training to enhance the overall performance.
We draw inspiration for this approach from~\cite{lee2014speaker}.
The output of the gating network is
\begin{equation}
    \bm{\alpha} = \mathcal{G}(\mathbf{e}_{\text{input}}),
\end{equation}
where $\bm{\alpha} \in \mathbb{R}^N$.
The final \gls{moe} output is obtained as in \eqref{eq:moe} and the final prediction score $\hat{y}$ is obtained likewise.


%% file: 3_setup.tex
\section{Experimental Setup}
\label{sec:setup}


\subsection{Speech Deepfake Detectors}

Our proposed method utilizes a \gls{moe} framework with four experts, each trained on a specific speech deepfake dataset.
By integrating these experts, we aim to leverage their specialized knowledge to develop a speech deepfake detector with enhanced detection capabilities for both seen and unseen deepfake datasets.

To ensure consistency across all our experiments, we used a single speech deepfake detection model, a LCNN model~\cite{wu2018light}, as both a baseline and expert within the \gls{moe} framework.
Despite its simplicity, this lightweight CNN-based model is renowned for its competitive and reliable performance, even when compared to more complex systems. For this reason, it has been considered as a baseline in the ASVspoof 2021 challenge~\cite{yamagishi2021asvspoof}.
Our version of the LCNN processes mel-spectrograms from the input audio. 
We trained four separate LCNN models on distinct datasets, with each model serving as an expert in our \gls{moe} framework.
Additionally, we trained a single LCNN model on a combination of all the training datasets.
We consider this model as a first baseline, as it benefits from exposure to all training datasets simultaneously, allowing it to better acquire knowledge from
and generalize across diverse data distributions.
As a second baseline, we consider the average ensemble as a simple, input-independent approach for combining scores from multiple experts~\cite{salvi2023robust}, making it a particularly suitable baseline to compare with the proposed \gls{moe} systems.

Each LCNN model processes \num{3} seconds of audio, corresponding to \num{48000} audio samples. 
Inputs shorter than this are padded with repeated data to meet the required length, while the longer ones are truncated.
We trained each individual expert for \num{100} epochs, monitoring the validation loss computed using the Cross-Entropy function.
We employed early stopping with a patience of \num{20} epochs, used a batch size of \num{128}, and applied the AdamW optimizer with an initial learning rate of \num{e-4}, adjusted according to a cosine annealing schedule. 
Each batch was balanced to include an equal number of samples from both the real and fake classes.

In both the standard \gls{moe} and enhanced \gls{moe}, the gating network architecture comprises a single fully-connected layer followed by dropout, batch normalization, and a Leaky ReLU activation function.  
At training time, 
all experts and the gating network were trained jointly, the former starting from the pre-trained weights, while the latter is randomly initialized.
To do so, we used the same setup as for individual experts, except the batch size was reduced to \num{64}.
The code used to train and test all the models is available online.\footnote{Code will be released upon acceptance.}

\subsection{Datasets}

During our experimental campaign, we employed six different speech deepfake detection datasets $D$. 
All audio data were sampled at a frequency of $f_\text{s} = \SI{16}{\kilo\hertz}$. \vspace{5pt}

\noindent \textbf{ASVspoof 2019~\cite{todisco2019asvspoof}} ($D_\text{ASV}$).
Released for the homonymous challenge, this dataset was designed to develop effective ASV models.
It includes real speech from the VCTK corpus~\cite{veaux2016superseded} and synthetic speech fabricated by \num{19} different synthetic speech generators. Data are distributed unevenly across the dataset partitions to enable open-set evaluation. We employ the Logical Access partition of this corpus.

\noindent \textbf{FakeOrReal~\cite{reimao2019dataset}} ($D_\text{FoR}$).
This dataset contains over \num{198000} utterances of both real and synthetic speech. The synthetic data were generated using various \gls{tts} systems, including both open-source and commercial tools. The real data are sourced from a diverse range of open-source speech datasets and additional sources, (TED Talks, YouTube, etc.).

\noindent \textbf{ADD 2022~\cite{yi2022add}} ($D_\text{ADD}$). This is a synthetic speech dataset released for the homonymous challenge. It contains real and synthetic speech data simulating realistic scenarios, such as background noises and disturbances. All the tracks contained in this set come from Chinese speakers. We include it in our experiments to introduce linguistic diversity, as all other datasets we consider consist of English speech.

\noindent \textbf{In-the-Wild~\cite{muller2022does}} ($D_\text{ItW}$). 
This dataset aims to evaluate speech deepfake detectors in realistic environments by providing in-the-wild speech data. It consists of nearly \num{38} hours of audio, evenly split between fake and real speech. The fake clips were created by segmenting publicly accessible video and audio files, while the real clips come from publicly available material featuring the same speakers.
We divided the dataset into three partitions: 60\% for training, 20\% for development, and 20\% for evaluation, ensuring a balanced distribution of real and fake samples in each split. 

\noindent \textbf{Purdue speech dataset~\cite{bhagtani2024recent}} ($D_\text{PUR}$).
This corpus includes \num{25000} synthetic speech tracks generated by five advanced diffusion model-based voice cloning methods: ProDiff, DiffGAN-TTS, ElevenLabs, UnitSpeech, and XTTS. The real speech data were sourced from LJSpeech~\cite{LJSpeech} and LibriSpeech~\cite{panayotov2015librispeech}.

\noindent \textbf{TIMIT-TTS~\cite{salvi2023timit}} ($D_\text{TIM}$).
This is a speech dataset that includes only fake audio samples, generated from 12 different TTS methods. It is created based on the VidTIMIT corpus~\cite{sanderson2002vidtimit} by generating a synthetic copy of its 430 tracks using all the considered synthetic speech generators. We consider the \textit{clean} partition of this corpus.

We used four of these datasets ($D_\text{ASV}$, $D_\text{FoR}$, $D_\text{ADD}$, $D_\text{ItW}$) for both training and testing the proposed models, while the remaining two ($D_\text{PUR}$, $D_\text{TIM}$) were employed only in the testing phase, to evaluate the detection capabilities of the considered systems on unseen data.

%% file: 4_results.tex
\section{Results}
\label{sec:results}

\begin{table*}[t]
\caption{EER (\%) and AUC (\%) values of the considered systems across known and unknown datasets. $\uparrow$ means the higher the better. $\downarrow$ means the lower the better.}
\label{tab:results}
\resizebox{\textwidth}{!}{
\begin{tabular}{lcccccccc|cccc|cccc}
\hline
\toprule
                        & \multicolumn{2}{c}{$D_\text{ASV}$} & \multicolumn{2}{c}{$D_\text{FoR}$} & \multicolumn{2}{c}{$D_\text{ADD}$} & \multicolumn{2}{c}{$D_\text{ItW}$} & \multicolumn{2}{c}{$D_\text{PUR}$} & \multicolumn{2}{c}{$D_\text{TIM}$} & \multicolumn{2}{c}{\textbf{Known}} & \multicolumn{2}{c}{\textbf{All}} \\ \midrule
\multicolumn{1}{l}{}    & EER $\downarrow$        & AUC $\uparrow$       & EER $\downarrow$        & AUC $\uparrow$       & EER $\downarrow$        & AUC $\uparrow$       & EER $\downarrow$       & AUC $\uparrow$       & EER $\downarrow$       & AUC $\uparrow$       & EER $\downarrow$       & AUC $\uparrow$       & EER $\downarrow$             & AUC $\uparrow$            & EER $\downarrow$           & AUC $\uparrow$            \\ \midrule \midrule
$\mathcal{E}_\text{ASV}$  & \textbf{8.8}        & \textbf{97.24}      & 23.10      & 86.33      & 47.48      & 53.26      & 43.56      & 59.07      & 33.45      & 72.29      & 11.40      & 94.40      & 30.74            & 73.98           & 27.97          & 77.10           \\
$\mathcal{E}_\text{FoR}$                     & 29.11      & 78.03      & 14.75      & 92.25      & 46.84      & 57.90      & 19.09      & 88.15      & 23.21      & 83.36      & 8.37       & 95.62      & 27.45            & 79.08           & 23.56          & 82.55           \\
$\mathcal{E}_\text{ADD}$                     & 55.62      & 36.39      & \textbf{2.34}       & 61.78      & 35.60      & 69.62      & 63.68      & 32.43      & 24.39      & 47.71      & 98.84      & 0.07       & 39.31            & 50.06           & 46.75          & 52.93           \\
$\mathcal{E}_\text{ItW}$                     & 24.23      & 81.76      & 19.30      & 89.72      & 43.55      & 59.31      & 0.38       & \textbf{99.89}      & 15.73      & 91.34      & 7.21       & 97.62      & 21.87            & 82.67           & 18.40          & 86.61           \\  \midrule
Ensemble                & 13.43      & 93.80      & 10.29      & 97.03      & 38.54      & 67.29      & 11.20      & 95.96      & 21.98      & 82.77      & 29.30      & 75.49      & 18.37            & 88.52           & 20.79          & 85.39           \\
Joint Training          & 9.29       & 96.57      & 3.80       & 98.56      & 30.86      & 72.23      & 0.98       & 99.73      & 12.38      & 84.66      & 3.26       & 94.26      & 11.23            & 91.77           & 10.10          & 91              \\
MoE (standard)          & 9.56       & 96.97      & 5.87       & 98.26      & \textbf{28.67}      & \textbf{76.20}      & \textbf{0.1}        & 99.66      & 10.98      & 84.28      & \textbf{0.93}       & 95.12      & 11.05            & 92.77           & 9.35           & 91.75           \\
\textbf{MoE (enhanced)} & 9.45       & 96.17      & 2.74       & \textbf{99.43}      & 30.87      & 76.04      & 0.55       & 99.83      & \textbf{2.53}       & \textbf{94.52}      & 6.98       & \textbf{97.73}      & \textbf{10.90}   & \textbf{92.87}  & \textbf{8.85}  & \textbf{93.95} \\ \bottomrule 
\end{tabular}}
\end{table*}


\subsection{Detection Performance}
\label{subsec:det_results}
As a first experiment, we evaluate the effectiveness of the proposed approach on the speech deepfake detection task.
We refer to the classic \gls{moe} as \textit{\gls{moe} (standard)} and to our customized \gls{moe} as \textit{\gls{moe} (enhanced)}, which utilizes the hidden representations of the experts.
We evaluate these systems against the performances of the four individual experts (denoted as $\mathcal{E}_\text{ASV}$, $\mathcal{E}_\text{FoR}$, $\mathcal{E}_\text{ADD}$ and $\mathcal{E}_\text{ItW}$), an averaging ensemble of such experts, and the LCNN model trained jointly at once on all the four training datasets.

The results are presented in \Cref{tab:results}, in terms of \gls{eer} and \gls{auc}.  
The ``Known'' column indicates the average performance on the evaluation partitions of the datasets $D_\text{ASV}$, $D_\text{FoR}$, $D_\text{ADD}$ and $D_\text{ItW}$, which were used during training, while the ``All'' column reflects the average performance across all the considered datasets, including $D_\text{PUR}$ and $D_\text{TIM}$.

The results reveal that the \gls{moe} models, particularly the \textit{\gls{moe} (enhanced)}, outperform other approaches in both \gls{eer} and \gls{auc}, demonstrating their effectiveness in handling known and unseen datasets. 
Specifically, the \textit{\gls{moe} (enhanced)} achieves the lowest \gls{eer}, with scores of $10.90\%$ on known datasets and $8.85\%$ across all domains.
The joint training strategy and the \textit{\gls{moe} (standard)} also perform well but are slightly less effective.
They achieve \gls{eer} scores of $11.23\%$ and $11.05\%$ on known datasets, and $10.10\%$ and $9.35\%$ on all sets, respectively.
The ensemble method delivers moderate performance, as it outperforms all the individual experts, but it is surpassed by both the joint training and \gls{moe} approaches.
On the other hand, the individual expert models ($\mathcal{E}_\text{ASV}$, $\mathcal{E}_\text{FoR}$, $\mathcal{E}_\text{ADD}$, and $\mathcal{E}_\text{ItW}$) show varied performance. 
Some experts excel on specific datasets but underperform on others.
Notably, the $\mathcal{E}_\text{ADD}$ expert exhibits inconsistent behavior, performing exceptionally well on $D_\text{FoR}$, but worse on $D_\text{ADD}$, its own training dataset.
This counterintuitive result is likely due to the dataset's construction, which involves disjoint deepfake algorithms between the training and testing partitions, presenting a significant challenge for the detectors.
In this regard, we observe that no system excels on $D_\text{ADD}$, all of them showing relatively poor results. 
This leads to an unusual situation where the average performance on the ``known'' datasets is often worse than on all datasets, as $D_\text{PUR}$ and $D_\text{TIM}$ seem not to be as challenging as $D_\text{ADD}$.

\begin{figure*}[t]
    \centering
    \begin{subfigure}[b]{0.32\textwidth}
        \includegraphics[width=\textwidth]{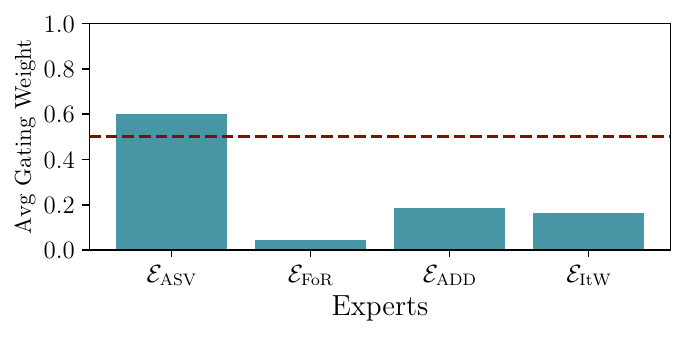}
        \caption{ASVspoof 2019 ($D_\text{ASV}$)}
    \end{subfigure}
    \hfill
    \begin{subfigure}[b]{0.32\textwidth}
        \includegraphics[width=\textwidth]{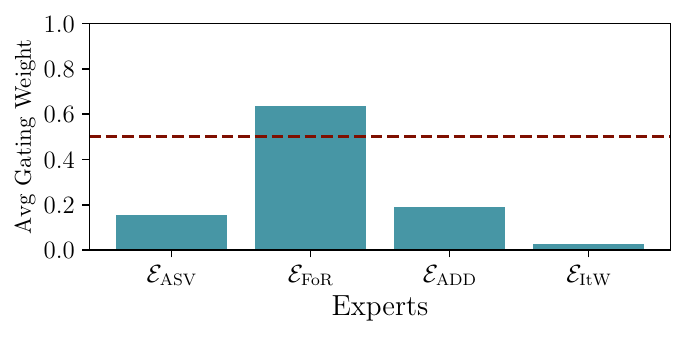}
        \caption{FakeOrReal ($D_\text{FoR}$)}
    \end{subfigure}
    \hfill
    \begin{subfigure}[b]{0.32\textwidth}
        \includegraphics[width=\textwidth]{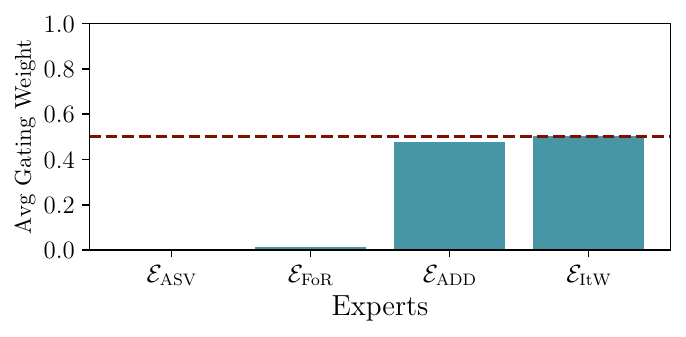}
        \caption{ADD 2022 ($D_\text{ADD}$)}
    \end{subfigure}

    \begin{subfigure}[b]{0.32\textwidth}
        \includegraphics[width=\textwidth]{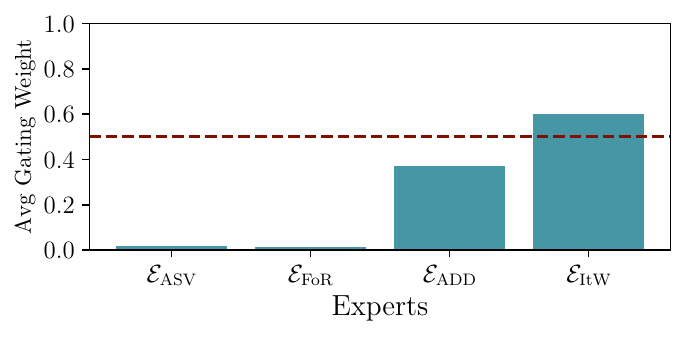}
        \caption{In-the-Wild ($D_\text{ItW}$)}
    \end{subfigure}
    \hfill
    \begin{subfigure}[b]{0.32\textwidth}
        \includegraphics[width=\textwidth]{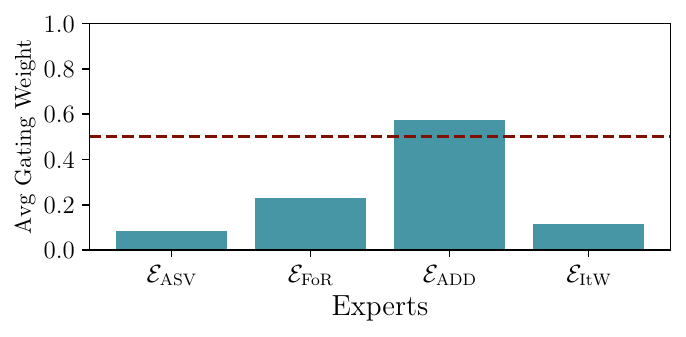}
        \caption{Purdue speech dataset ($D_\text{PUR}$)}
    \end{subfigure}
    \hfill
    \begin{subfigure}[b]{0.32\textwidth}
        \includegraphics[width=\textwidth]{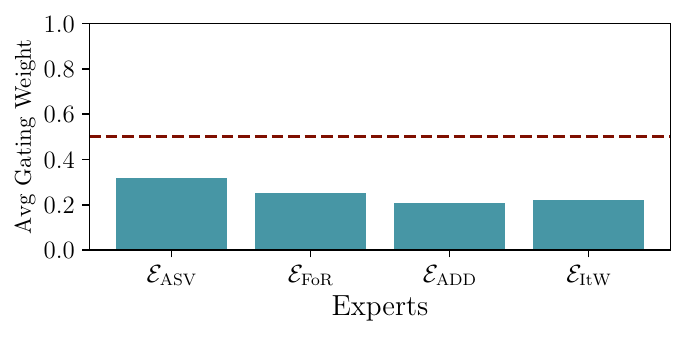}
        \caption{TIMIT-TTS ($D_\text{TIM}$)}
    \end{subfigure}
    
    \caption{Average gating weights distributions for all the considered datasets. The weights represent the proportion of each expert's contribution to the decision-making process.}
    \label{fig:gating_weights}
\end{figure*}

\subsection{Gating Network Analysis}
\label{subsec:gate_results}
We now perform a further analysis of the gating network in the best-performing system, i.e., \textit{\gls{moe} (enhanced)}.
By examining the average gating weights across different datasets, we aim to evaluate several aspects: 
(1) whether the gate selects the expert for which pre-training and test dataset coincide,
(2) the selectiveness of the gating weights' softmax distribution, 
(3) the relative contribution of each expert to predictions on speech samples from unknown datasets.

\Cref{fig:gating_weights} shows the results of this analysis for each individual dataset. 
For the datasets $D_\text{ASV}$, $D_\text{FoR}$, and $D_\text{ItW}$, the expert with the highest average weight is the one specifically pre-trained on that dataset. This was largely expected since the corresponding experts showcase high performance on the respective datasets. 
Conversely, for $D_\text{ADD}$, the experts $\mathcal{E}_\text{ADD}$ and $\mathcal{E}_\text{ItW}$ show nearly equal importance. This observation aligns with the results in \Cref{tab:results}, where $\mathcal{E}_\text{ADD}$ performs better than other experts within its own domain but does not excel to the same extent as they do within theirs.
This likely explains why the gating network assigns substantial weight to $\mathcal{E}_\text{ItW}$, which performed second-best on $D_\text{ADD}$.
The analysis results for the two unknown datasets, $D_\text{PUR}$ and $D_\text{TIM}$, provide intriguing insights. 
For $D_\text{PUR}$, which includes fake clips generated by advanced diffusion models, the expert trained on $D_\text{ADD}$ emerges as the most relevant. 
In contrast, for $D_\text{TIM}$, all experts contribute almost equally to the final predictions.
These results underscore the effectiveness of a \gls{moe}-based approach and a gating network system in effectively integrating multiple experts for the task at hand. 
Moreover, it is noteworthy how the gating network can also serve as a valuable analysis tool, offering insights into the similarity between different domains, enabling us to better understand the relationship between them. 


%% file: 5_conclusion.tex
\section{Conclusions}
\label{sec:conclusion}

In this paper, we present a novel speech deepfake detection method based on the \gls{moe} framework. 
Our approach leverages the divide-and-conquer strategy inherent to \glspl{moe}, aiming to address the generalization challenges faced by current deepfake detectors.
Extensive evaluations across multiple state-of-the-art datasets demonstrate the effectiveness of our approach in managing data from diverse domains. 
As this is an initial exploration of applying \glspl{moe} to speech deepfake detection, future work will focus on experimenting with new \gls{moe} architectures and learning strategies for the gating network.
Additionally, we plan to assess the scalability of our model by increasing the number of experts.